\documentclass[aps,prc,twocolumn,showpacs,groupedaddress]{revtex4}

\usepackage{amstext}

\usepackage{graphicx}
\usepackage{amsmath}
\usepackage{amssymb}
\usepackage{dcolumn}
\usepackage{bm}

\begin{document}

\title{Peculiarity of the stellar CNO cycle with energetic particles}

\author{V.~T.~Voronchev}
\affiliation{Skobeltsyn Institute of Nuclear Physics, Lomonosov Moscow State University, Moscow 119991, Russia}
\email[]{voronchev@srd.sinp.msu.ru}


\begin{abstract}
A peculiarity of the stellar CNO cycle caused by MeV $\alpha$-particles and protons generated in exoergic nuclear processes is analyzed. The main parameters of these particles and suprathermal reactions induced by them in a stellar core are calculated. It is shown that these reactions can trigger an abnormal nuclear flow in the second branch of the stellar CNO cycle. A conjecture is made that the phenomenon is of a general nature and can manifest in various stars at non-exploding stages of their evolution. The influence of the abnormal flow on some CNO characteristics is demonstrated.
\end{abstract}


\maketitle

\section{Introduction}
\label{intro}

It is known that most models of stellar nucleosynthesis rely on nuclear reaction networks operating with thermal processes between Maxwellian particles. In some cases, however, energetic non-Maxwellian particles of various origins are also invoked to account for element abundances in certain stars (see, e.g., Ref.~\cite{arno20} and references therein). For example, this concerns characteristics of the surface composition of P-rich stars \cite{gori22}, metal-poor halo stars \cite{tati18}, and a chemically-peculiar magnetic star \cite{gori07}. The nuclear interaction of solar energetic particles with the surface of the early Sun is also considered \cite{liu12} as a mechanism for the origin of some short-lived nuclides.

Characteristics of the inner part of stars can also be influenced by energetic particles naturally produced in thermonuclear reactions and released in the decay of unstable nuclei. These fast projectiles slow down in stellar core plasmas and with some probability undergo suprathermal nuclear reactions before being thermalized. While these reactions are omitted in thermal reaction networks, they can significantly enhance endoergic processes that do not proceed at low energies below their thresholds. A possibility of such effects in some stellar reactions induced by fast charged particles and neutrons was demonstrated, e.g., in Refs.~\cite{beau77,petr88,shap04}. The reaction enhancement can particularly appear in reverse and breakup processes, as they are all endoergic (for breakup reactions in the Sun, see Ref.~\cite{voro21a}).

In this context, the stellar CNO cycle presents a sequence of processes suitable for the suprathermal effects to manifest. This cycle involves various reverse processes like $\alpha + \mathrm{B} \rightarrow p + \mathrm{A}$ that can be induced by fast $\alpha$-particles generated in stellar reactions.

One of interesting processes here is
\begin{equation}\label{eq:rev}
    \alpha + \mathrm{^{14}N} \rightarrow p+\mathrm{^{17}O} \quad (E_{\alpha,\text{thr}} = 1.531\text{ MeV})
\end{equation}
having a moderate value of threshold $E_{\alpha ,\text{thr}}$ accessible for MeV $\alpha$-particles. The balance between this process and the corresponding forward $(p,\alpha )$ reaction
\begin{equation}\label{eq:for}
    p+\mathrm{^{17}O} \rightarrow \alpha + \mathrm{^{14}N} \quad (Q = 1.191\text{ MeV})
\end{equation}
controls nuclear flow in the $\mathrm{^{14}N} - \mathrm{^{17}O}$ pair. An important point is that the $(p,\alpha )$ reaction closes the second branch of the CNO cycle (the CNO-II cycle) and is one of three processes primarily determining the CNO burning rate \cite{adel98}.

Some features of the solar CNO cycle irradiated by MeV $\alpha$-particles born in the $pp$ chain processes were demonstrated in Refs.~\cite{voro17a,voro17b}. In particular, it was shown that the reverse $\mathrm{^{14}N}(\alpha ,p)\mathrm{^{17}O}$ reaction rate can exceed the forward $\mathrm{^{17}O}(p,\alpha )\mathrm{^{14}N}$ one, causing the redirection of nuclear flow between $\mathrm{^{14}N}$ and $\mathrm{^{17}O}$. It was also obtained that this phenomenon can affect some characteristics of the CNO cycle, e.g., the $^{17}$O abundance and the emission rate of $^{18}$F neutrinos in the solar outer core \cite{voro19}.

In view of these findings, an interesting question arises whether the CNO-II flow peculiarity marked for the Sun is of a general nature that creates conditions for its manifestation in other stars at non-exploding stages of their evolution.

The purpose of the present report is to provide arguments in favor of such a scenario.

\section{Abnormal nuclear flow and its effect on CNO parameters}
\label{cno}

Several factors determine the rates of suprathermal reactions in a stellar plasma. One of them is the flux of fast particles $f$ generated in the stellar core. In the Sun, most of these particles are produced in the $pp$ chain processes being the predominant mechanism of solar burning. Given this, in order to explore the question posed in Sec.~\ref{intro}, it seems reasonable to examine suprathermal processes in objects other than the Sun, in which not the $pp$ chain but the CNO cycle plays an important role in stellar burning.

Such an object -- a star with a mass $M = 5M_\odot$ and a metallicity $Z = 0.02$ having radial temperature and density profiles found \cite{ayukov} by running a code \cite{mesa} -- is chosen for the present analysis. It is worth noting that the plasma condition in this star essentially differs from that in the Sun. For example, in the star's center the plasma temperature $T = 2.5$~keV (versus the solar core temperature $T_\odot \sim 1.4$~keV) and density $\rho = 20$~g/cm$^3$ (versus $\rho_\odot \sim 150$~g/cm$^3$).

\subsection{Calculation model}
\label{model}

The formalism of in-flight reaction probability was used to calculate the rate $R_{fb \rightarrow xy,\text{sth}}$ of a suprathermal $f + b \rightarrow x +y$ reaction induced by fast charged particle $f$ in the stellar core plasma. According to it
\begin{equation}\label{eq:Rsth}
    R_{fb \rightarrow xy,\text{sth}} =
    R_f \times W_{fb \rightarrow xy},
\end{equation}
where $R_f$ is the emission rate of particles $f$ and $W_{fb \rightarrow xy}$ is the probability that they undergo the in-flight $f+b$ reaction with bulk ions $b$ while slowing down in the plasma. The particle emission rate in a $i+j$ process is
\begin{equation}\label{eq:Rf1}
    R_f =
    N_f (1+\delta_{ij})^{-1} n_in_j \langle\sigma v\rangle_{ij},
\end{equation}
where $N_f$ is the number of particles $f$ produced per pair of $(ij)$, $n_i$ and $n_j$ are the number densities of plasma species $i$ and $j$, and $\langle\sigma v\rangle_{ij}$ is the thermal (that is, Maxwellian) $i+j$ reactivity. If fast particles are released in the decay of an unstable nucleus $Y$, then
\begin{equation}\label{eq:Rf2}
    R_f = N_f n_Y/\tau,
\end{equation}
where $\tau$ is the nucleus mean lifetime.

The expression for $W_{fb \rightarrow xy}$ has a different form for monoenergetic particles $f$ and those with some source energy spectrum. In the former case
\begin{eqnarray}\label{eq:Wfb1}
    W_{fb \rightarrow xy}(E_{f,0} \rightarrow E_{f,1}) =
      & 1 &
    - \exp
     \left [
    \int\limits_{E_{f,1}}^{E_{f,0}}
    \left(\frac{2E_f}{m_f}\right)^{1/2}
     \right. \nonumber \\
      & \times &
     \left.
    n_b \frac{\sigma_{fb \rightarrow xy} (E_f)}{(dE_f/dt)}
    \, dE_f
     \right ].
\end{eqnarray}
In Eq.~(\ref{eq:Wfb1}), $E_{f,0}$ and $E_{f,1}$ are the initial and final particle energy, respectively, $\sigma_{fb \rightarrow xy}$ is the reaction cross section, and $(dE_f/dt)$ is the particle energy loss rate in the plasma. In turn, for fast particles having a source energy spectrum  $S(E'_f)$ with $0 \leq E'_f \leq E_{f,\text{max}}$
\begin{equation}\label{eq:Wfb2}
    W_{fb \rightarrow xy} =
    \frac{\int\limits_{E_{f,1}}^{E_{f,\text{max}}}
    W_{fb \rightarrow xy}(E'_f \rightarrow E_{f,1})
    S(E'_f) \,dE'_f}
    {\int\limits_0^{E_{f,\text{max}}} S(E'_f) \,dE'_f}.
\end{equation}

One of the key parameters determining the probability $W_{fb \rightarrow xy}$ is the particle energy loss rate $(dE_f/dt)$. In the stellar plasma, most of $(dE_f/dt)$ comes from Coulomb (Coul) scattering of particles $f$ by bulk electrons ($e$) and ions ($i$), that is
\begin{equation}\label{eq:eloss}
    (dE_f/dt) \simeq (dE_f/dt)^\text{Coul}_e + (dE_f/dt)^\text{Coul}_i.
\end{equation}
The terms $(dE_f/dt)^\text{Coul}_s$ ($s=e,i$) can be described using a model \cite{kame86} based on the Fokker-Planck collision theory. They are given by
\begin{eqnarray}\label{eq:coul}
    \left( \frac{dE_f}{dt} \right)_s^\text{Coul} =
      & - &
    \frac{8\pi^2 (Z_f Z_s)^2 e^4 (2m_f)^{1/2}}{m_s E^{1/2}_f}
    \ln\Lambda_{f s} \nonumber \\
      & \times &
    J(v_f)
\end{eqnarray}
with
\begin{eqnarray}\label{eq:J}
    J(v_f)
      & = &
    \int^{v_f}_0
    v^2_s f_s(v_s) \,dv_s
    -\frac{m_s}{3E_f}
     \left (
    \int^{v_f}_0 v^4_s f_s(v_s) \,dv_s
     \right. \nonumber \\
      & + &
     \left.
    v^3_f \int^\infty_{v_f} v_s f_s(v_s) \,dv_s
     \right ).
\end{eqnarray}
In Eqs.~(\ref{eq:coul}) and (\ref{eq:J}), $\ln\Lambda_{p s}$ is the Coulomb logarithm and $f_s(v_s)$ is the density-normalized velocity distribution function of species $s$.

\subsection{Numerical results}
\label{result}

\begin{table}
\caption{\label{tab:reactions}A list of processes generating fast particles $f (= p,\alpha )$ in the stellar core. The right column shows the particle energy $E_f$. For the process No.~3, the $\alpha$-particle energy may change within the half-width of the $\mathrm{^8Be^\ast[16.626]}$ state \cite{till04}.}
\begin{ruledtabular}
\begin{tabular}{ccc}
No. & Reaction & $E_f$ (MeV) \\
\hline
1 & $\mathrm{^3He}(\mathrm{^3He},2p)\alpha$ & $\lesssim 10.7$ (for $p$)     \\
  &                                         & $\lesssim 4.3$ (for $\alpha$) \\
2 & $\mathrm{^7Li}(p,\alpha)\alpha$ & 8.674 \\
3 & $\mathrm{^8B}(\beta^+)\mathrm{^8Be^\ast [16.626]}\rightarrow 2\alpha$ & 8.359 \\
4 & $\mathrm{^{15}N}(p,\alpha)\mathrm{^{12}C}$ & 3.724 \\
5 & $\mathrm{^{18}O}(p,\alpha)\mathrm{^{15}N}$ & 3.142 \\
\end{tabular}
\end{ruledtabular}
\end{table}

\begin{figure}
\begin{center}
\includegraphics[width=7.5cm]{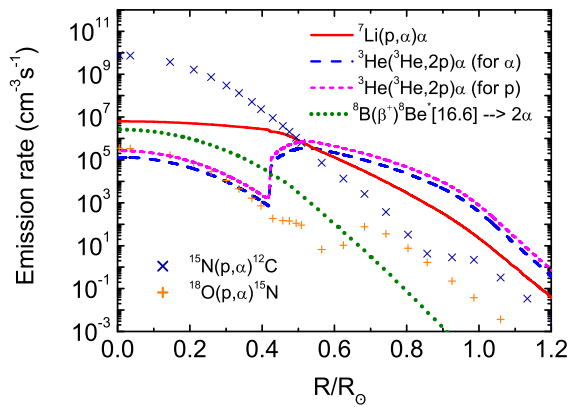}
\caption{\label{fig:emission} The emission rates of fast protons $R_p$ and $\alpha$-particles $R_\alpha $ as a function of distance $R/R_\odot$ from the center of the star.}
\end{center}
\end{figure}

The main processes generating fast particles -- protons and $\alpha$-particles -- in the stellar core are listed in Table~\ref{tab:reactions}. The particle emission rates $R_f$ in these processes (see Eqs.~(\ref{eq:Rf1}) and (\ref{eq:Rf2})) at different radii $R/R_\odot$ are shown in Fig.~\ref{fig:emission}. The thermal reactivities $\langle\sigma v\rangle_{ij}$ used to calculate these rates were taken from compilations \cite{xu13,ilia10}. It is seen that the maximum value $R_\alpha ^\text{max}$ reaches $8\times 10^9$~cm$^{-3}$s$^{-1}$ and is provided by the CNO reaction $\mathrm{^{15}N}(p,\alpha)\mathrm{^{12}C}$. It is worth noting that in the Sun $R_\alpha ^\text{max}$ is less by two orders of magnitude and is provided by the $pp$ chain process $\mathrm{^7Li}(p,\alpha)\alpha$ \cite{voro17b}.

\begin{figure}
\begin{center}
\includegraphics[width=7.5cm]{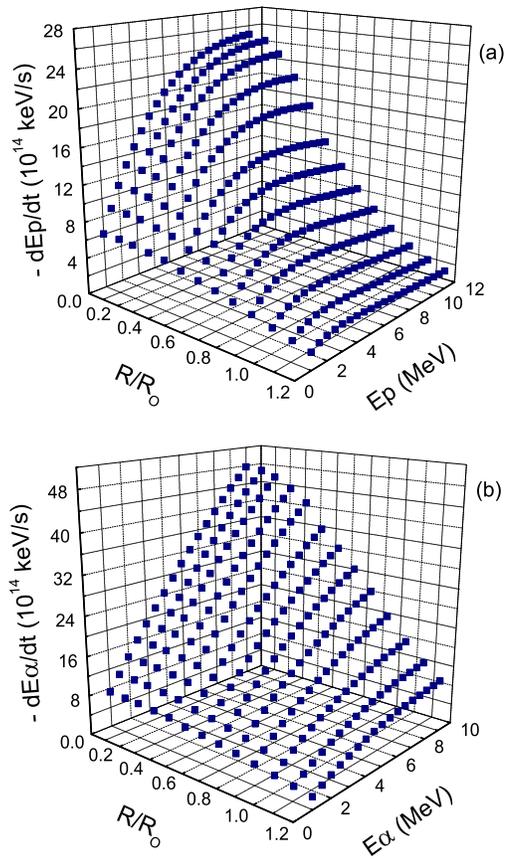}
\caption{\label{fig:loss} The energy loss rates of fast protons (a) and $\alpha$-particles (b) in different regions of the stellar core.}
\end{center}
\end{figure}

The calculated proton and $\alpha$-particle energy loss rates $dE_p/dt$ and $dE_\alpha/dt$, Eqs.~(\ref{eq:eloss})--(\ref{eq:J}), are plotted in Figs.~\ref{fig:loss}(a) and \ref{fig:loss}(b). One should note here that for the considered energy $E_\alpha \leq 8.674$~MeV (see Table~\ref{tab:reactions}) the $\alpha$-particle velocity $v_\alpha$ does not exceed the average velocity $\langle v_e \rangle$ of plasma electrons, which have the temperature in the range of 1--2.5~keV. As for 10.7-MeV protons, their velocity $v_p$ is higher than $\langle v_e \rangle$ by a factor of 2. Although this may introduce some uncertainty in the value of $dE_p/dt$, in the present case it does not matter much because the role of these protons is insignificant compared to that of the fast $\alpha$-particles.

It is found that the thermalization range $l_\text{th}(E_\text{initial} \rightarrow 3T/2)$ and time $\tau_\text{th}(E_\text{initial} \rightarrow 3T/2)$ of all these particles in the plasma are at most 0.1~cm and 60~ps, respectively. Despite the rapid thermalization, however, they are able to undergo suprathermal reactions with bulk ions.

\begin{figure}
\begin{center}
\includegraphics[width=7.5cm]{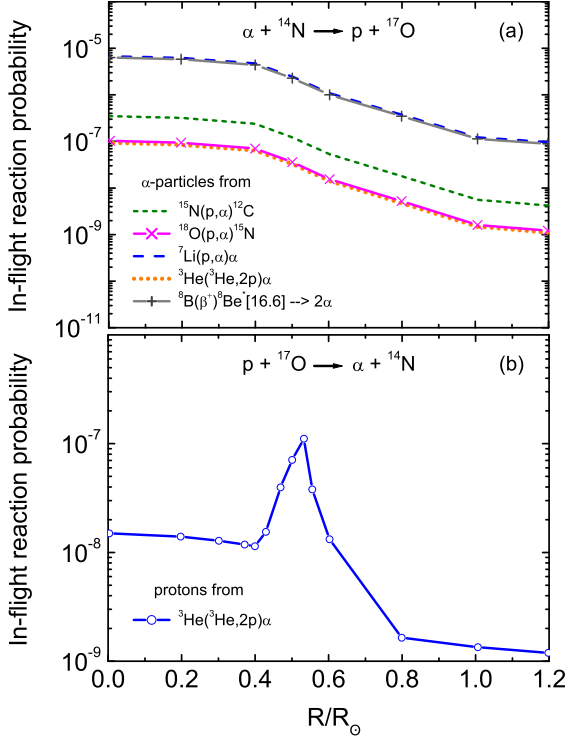}
\caption{\label{fig:prob} The probabilities of the reverse (a) and forward (b) processes induced by fast $\alpha$-particles and protons.}
\end{center}
\end{figure}

The probabilities $W_{fb \rightarrow xy}$, Eqs.~(\ref{eq:Wfb1}) and (\ref{eq:Wfb2}), of the reverse $\mathrm{^{14}N}(\alpha ,p)\mathrm{^{17}O}$ and forward $\mathrm{^{17}O}(p,\alpha )\mathrm{^{14}N}$ processes for fast particles in Table~\ref{tab:reactions} are presented in Figs.~\ref{fig:prob}(a) and \ref{fig:prob}(b). The results for the non-monoenergetic $^3$He+$^3$He protons and $\alpha$-particles were found with the source energy spectra $S$ in Eq.~(\ref{eq:Wfb2}) obtained in an $R$-matrix analysis \cite{brun15}.

\begin{figure}
\begin{center}
\includegraphics[width=7.5cm]{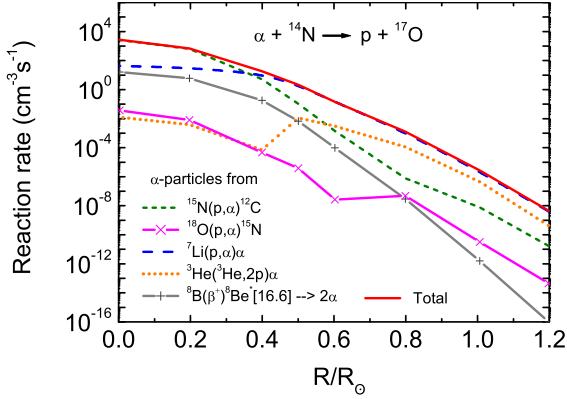}
\caption{\label{fig:a-rates} The total suprathermal $\mathrm{^{14}N}(\alpha ,p)\mathrm{^{17}O}$ rate together with partial contributions provided by fast $\alpha$-particles born in different processes.}
\end{center}
\end{figure}

The rate $R_{\alpha \mathrm{^{14}N} \rightarrow p\mathrm{^{17}O},\text{sth}}$ of the suprathermal $\mathrm{^{14}N}(\alpha ,p)\mathrm{^{17}O}$ reaction corresponding to the probability in Fig.~\ref{fig:prob}(a) and the $\alpha$-particle emission rate in Fig.~\ref{fig:emission} is given in Fig.~\ref{fig:a-rates}. Shown are the individual rates provided by various $\alpha$-particles together with the total rate. As for the thermal reaction rate $R_{\alpha \mathrm{^{14}N} \rightarrow p\mathrm{^{17}O},\text{th}}$, it is fully negligible. Indeed, it follows from the relation between thermal reverse and forward reactivities \cite{fowl67} that
\begin{eqnarray}\label{rev-for}
    R_{\alpha \mathrm{^{14}N} \rightarrow p\mathrm{^{17}O},\text{th}}
     & \simeq &
    0.7 \frac{n_\mathrm{^{4}He} n_\mathrm{^{14}N}}{n_\mathrm{^{1}H} n_\mathrm{^{17}O}}
    R_{p\mathrm{^{17}O} \rightarrow \alpha \mathrm{^{14}N},\text{th}} \nonumber \\
     & \times &
    \exp(-1191/T).
\end{eqnarray}
The temperature $T$ in the stellar core varies within 0.9--2.5~keV and, accordingly, $R_{\alpha \mathrm{^{14}N} \rightarrow p\mathrm{^{17}O},\text{th}} \simeq 0$.

\begin{figure}
\begin{center}
\includegraphics[width=7.5cm]{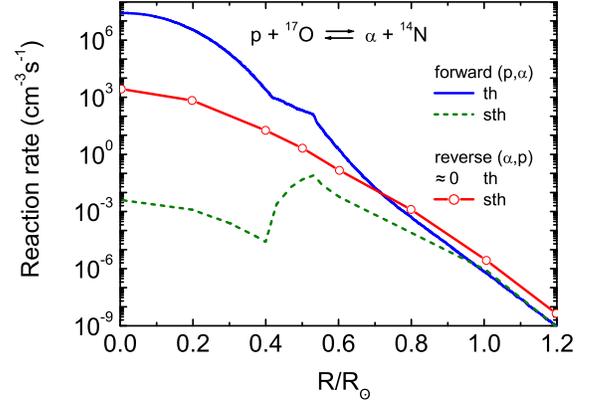}
\caption{\label{fig:rates} The thermal (th) and suprathermal (sth) rates for the forward $\mathrm{^{17}O}(p,\alpha )\mathrm{^{14}N}$ and reverse $\mathrm{^{14}N}(\alpha ,p)\mathrm{^{17}O}$ reactions in the stellar plasma.}
\end{center}
\end{figure}

\begin{figure}
\begin{center}
\includegraphics[width=7.5cm]{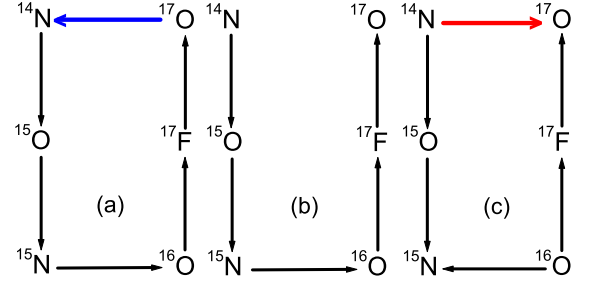}
\caption{\label{fig:cno} The diagram of the CNO-II cycle (see explanations in the text).}
\end{center}
\end{figure}

The total reverse $\mathrm{^{14}N}(\alpha ,p)\mathrm{^{17}O}$ rate is compared with the forward $\mathrm{^{17}O}(p,\alpha )\mathrm{^{14}N}$ rate in Fig.~\ref{fig:rates}. Both thermal and suprathermal contributions to these rates are shown. The results were found with the $\mathrm{^{14}N}(\alpha ,p)\mathrm{^{17}O}$ and $\mathrm{^{17}O}(p,\alpha )\mathrm{^{14}N}$ cross sections \cite{koni15,kies79,terw08} and the thermal reactivity $\langle\sigma v\rangle_{p\mathrm{^{17}O} \rightarrow \alpha\mathrm{^{14}N}}$ taken from Ref.~\cite{ilia10}.

The balance of these rates determines the nuclear flow between $^{14}$N and $^{17}$O. Figure~\ref{fig:rates} shows that in the outer core region the reverse $(\alpha ,p)$ rate is twice as high as the forward $(p,\alpha )$ one. This causes the flow redirection in the CNO-II cycle illustrated in Figs.~\ref{fig:cno}(a)--\ref{fig:cno}(c). The normal flow $\mathrm{^{14}N} \leftarrow \mathrm{^{17}O}$ (Fig.~\ref{fig:cno}(a)) is observed in the inner core. However, at the radius $R/R_\odot \gtrsim 0.7$ the situation changes. At $R/R_\odot \simeq 0.7$ the competing $(p,\alpha )$ and $(\alpha ,p)$ rates become equal (that is schematically demonstrated in Fig.~\ref{fig:cno}(b)) and at the larger radii the abnormal clockwise flow $\mathrm{^{14}N} \rightarrow \mathrm{^{17}O}$ appears (Fig.~\ref{fig:cno}(c)). This flow is formed in the outer core region constituting $\sim 70$\% of the total core volume.

\begin{figure}
\begin{center}
\includegraphics[width=7.5cm]{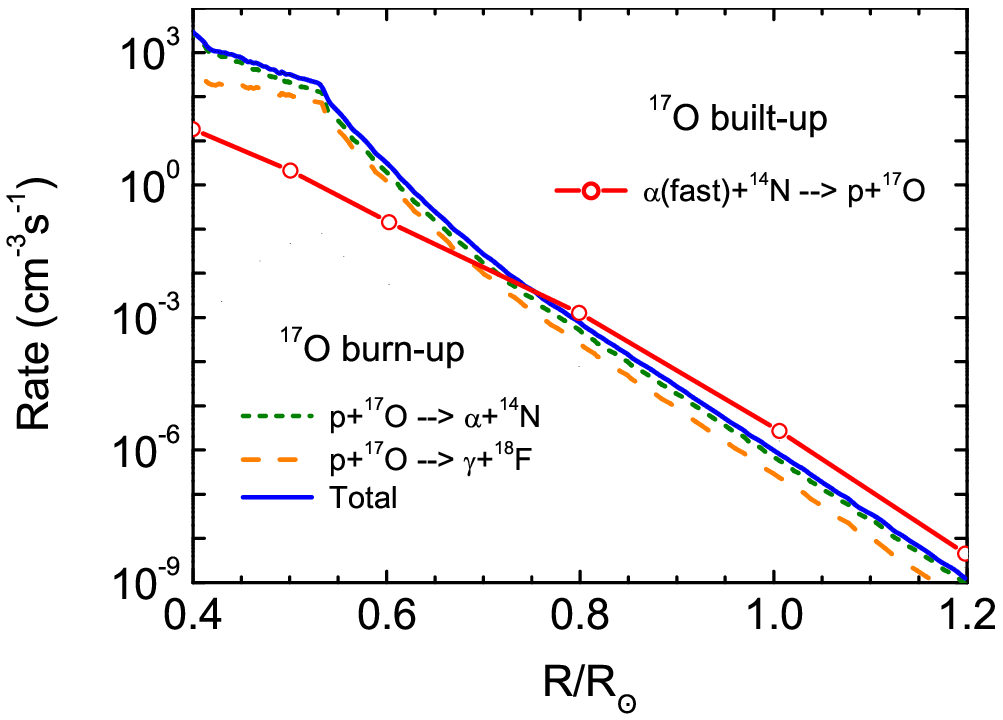}
\caption{\label{fig:17O} The rates of competing reactions of $^{17}$O built-up and burn-up in the stellar core.}
\end{center}
\end{figure}

To better realize the level of the abnormal flow as a channel of $^{17}$O built-up, it is useful to compare its rate with the rate of $^{17}$O burn-up. It is known that $^{17}$O burns up predominantly in $\mathrm{^{17}O}(p,\alpha )\mathrm{^{14}N}$ and $\mathrm{^{17}O}(p,\gamma )\mathrm{^{18}F}$ reactions. Their rates are compared with the $\mathrm{^{14}N}(\alpha ,p)\mathrm{^{17}O}$ rate in Fig.~\ref{fig:17O}. It is seen that in the outer core the suprathermal channel of $^{17}$O synthesis fully compensates both $^{17}$O burn-up
processes.

\begin{figure}
\begin{center}
\includegraphics[width=7.5cm]{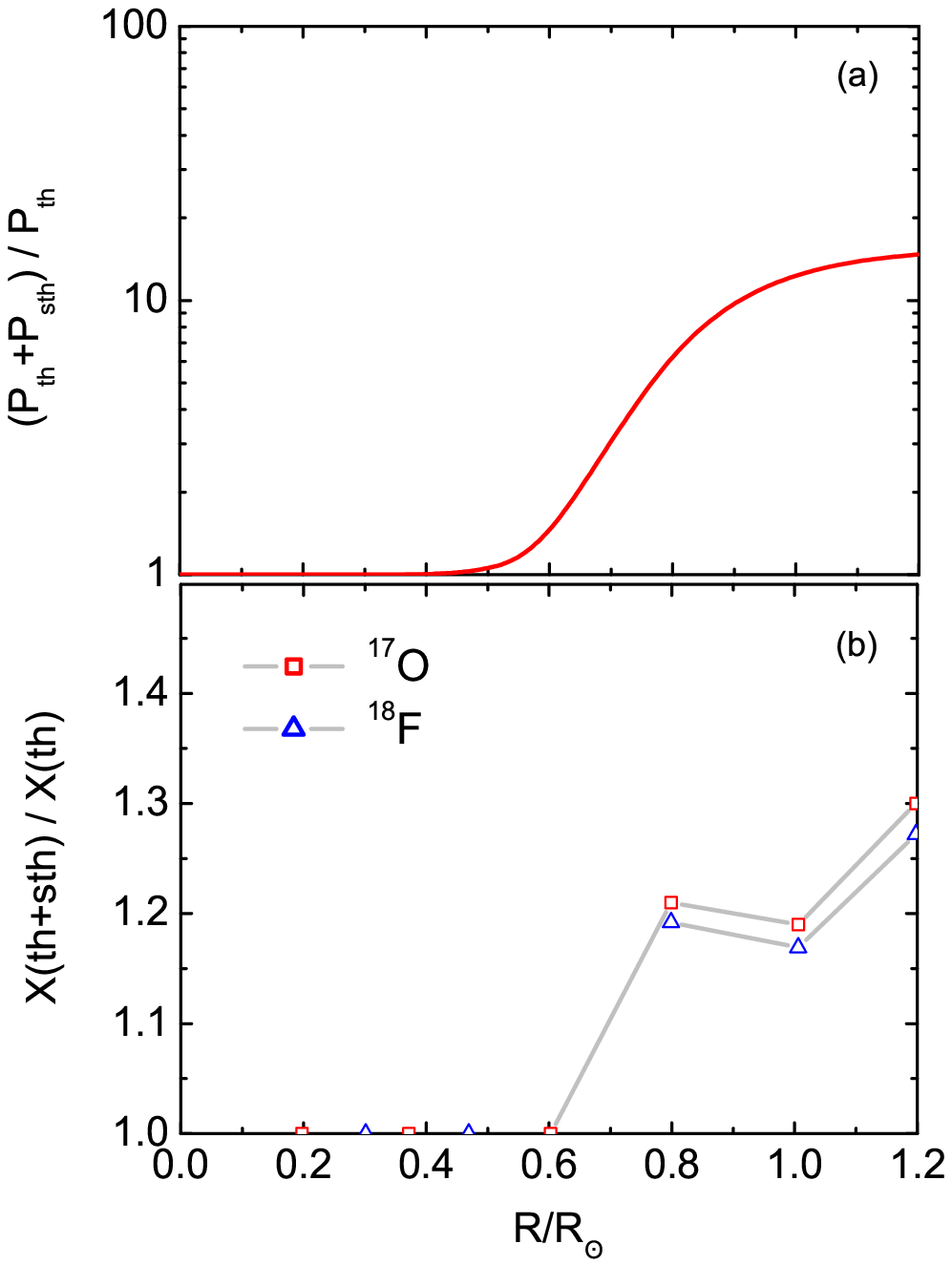}
\caption{\label{fig:enhanc} The suprathermal enhancement of some parameters related to the CNO cycle. (a). The nuclear power density for the $\mathrm{^{17}O}+p \rightleftarrows \alpha +\mathrm{^{14}N}$ processes. (b). The mass fractions of $^{17}$O and $^{18}$F. The latter curve also shows the enhancement of the $^{18}$F neutrino emission rate.}
\end{center}
\end{figure}

The results in Figs.~\ref{fig:rates} and \ref{fig:17O} suggest that the suprathermal reactions can affect the total energy released in the pair of $\mathrm{^{17}O}+p \rightleftarrows \alpha +\mathrm{^{14}N}$ processes and also change the $^{17}$O abundance in the outer core. The total power density $P$ for these processes is determined by a superposition of thermal and suprathermal components
\begin{equation}\label{eq:P}
    P = P_\text{th} + P_\text{sth},
\end{equation}
where
\begin{equation}\label{eq:Pth}
    P_\text{th} =
    R_{p\mathrm{^{17}O} \rightarrow \alpha \mathrm{^{14}N},\text{th}} \times Q,
\end{equation}
\begin{equation}\label{eq:Psth}
    P_\text{sth} =
    R_{f} \times \mathcal{E}.
\end{equation}
For an expression for the energy $\mathcal{E}$ released per fast particle $f(=\alpha ,p)$, the reader is referred to Ref.~\cite{voro19}. The calculated power densities are compared in Fig.~\ref{fig:enhanc}(a). One can see that at $R/R_\odot \geq 0.5$ the suprathermal component increases the total power density up to 15 times.

Figure~\ref{fig:enhanc}(b) presents the suprathermal enhancement for the mass fraction $X$ of $^{17}$O and also $^{18}$F that follows $^{17}$O in the CNO cycle diagram ( see, e.g., Ref.~\cite{wies10}). It is seen that in the outer core the suprathermal reactions increase the abundance of these elements by about 10\%--30\%. Furthermore, since the $^{18}$F decay rate $R_\nu = n_\mathrm{^{18}F} / \tau$, the curve marked with triangles in Fig.~\ref{fig:enhanc}(b) also shows the enhancement of $^{18}$F neutrino emission. One should note, however, that at $R/R_\odot > 1.2$ the suprathermal effect was found to be rapidly weakening. The results in Fig.~\ref{fig:enhanc}(b) were obtained by running a stellar reaction network allowing for the suprathermal processes, using a code \cite{code}. The code was run at constant temperature and density corresponding to the selected radii $R/R_\odot$.

\begin{table*}
\caption{\label{tab:comparison}A comparison of some plasma characteristics and suprathermal impact factors for the present object with those for the Sun \cite{voro19}. The parameters are shown for the inner core-outer core region.}
\begin{ruledtabular}
\begin{tabular}{lcc}
 & \multicolumn{2}{c}{Object} \\
 \cline{2-3}
 & Present & Sun \\
\hline
\multicolumn{3}{c}{Plasma characteristics} \\
predominant burning mechanism  & the CNO cycle & the $pp$ chain \\
temperature $T_e = T_i = T$ (keV) & 2.5--0.9 & 1.4--0.7 \\
density $\rho$ (g/cm$^3$) & 20--2 & 154--21 \\
electron number density $n_e$ (cm$^{-3}$) & 8$\times$$10^{24}$--9$\times$10$^{23}$ & 6$\times$$10^{25}$--1$\times$10$^{25}$ \\
particle emission rate $R_{f (=\alpha ,p)}$ (cm$^{-3}$s$^{-1}$) & 8$\times$$10^{9}$--6$\times$$10^{2}$ & 6$\times$$10^{7}$--3$\times$$10^{5}$ \\
degeneracy parameter $\Theta$\footnotemark[1] & 17--26 & 2--4 \\
coupling parameter $\Gamma$\footnotemark[2] & 0.02--0.03 & 0.07--0.07 \\
 & & \\
\multicolumn{3}{c}{Suprathermal impact} \\
nuclear flow redirection point $R_{r}/R_\odot$ & 0.7 & 0.08 \\
enhancement factor: & & \\
$\mathrm{^{17}O}+p \rightleftarrows \alpha +\mathrm{^{14}N}$ power density & 1--15 & 1--254 \\
$^{17}$O mass fraction & 1--1.3 & 1--4.1 \\
$^{18}$F mass fraction & 1--1.27 & 1--4.2 \\
$^{18}$F neutrino emission rate & 1--1.27 & 1--4.2 \\
\end{tabular}
\end{ruledtabular}
\footnotetext[1]{$\Theta = T_e / E_F$, where the Fermi energy $E_F = (\hbar^2/2m_e) (3\pi^2 n_e)^{2/3}$.}
\footnotetext[2]{$\Gamma = E_C / T_e$, where the Coulomb energy $E_C = e^2 (4\pi \varepsilon_0)^{-1} (4\pi n_e/3)^{1/3}$.}
\end{table*}

\section{Main conclusions}
\label{conc}

Thus, the irradiation of the stellar core plasma by reaction-produced MeV particles gives rise to the appearance of the abnormal nuclear flow between $^{14}$N and $^{17}$O capable of enhancing some CNO parameters in the outer core. These are the energy released in the $\mathrm{^{17}O}+p \rightleftarrows \alpha +\mathrm{^{14}N}$ reactions closing the CNO-II cycle, the abundances of $^{17}$O and $^{18}$F isotopes, and the $^{18}$F neutrino emission rate.

Some characteristics of the plasma are reduced in Table~\ref{tab:comparison}. Shown are the predominant mechanism of plasma burning and several parameters for the inner and outer core region. The respective characteristics for the Sun are also given for comparison. Table~\ref{tab:comparison} indicates that the physical conditions of these plasmas differ significantly. In particular, this concerns the parameters $R_f$ and $n_e$\footnote{Plasma electrons are the main stopping agents for fast particles $f$.} essentially affecting the strength of the suprathermal processes.

At the same time, however, a comparison of the present results with those for the Sun (see Table~\ref{tab:comparison}) shows that both objects demonstrate a qualitatively similar picture of suprathermal effects, such as the nuclear flow redirection and the parameter enhancement. This serves as an argument in favor of a conjecture that the phenomenon is of a general nature and can manifest to a greater or lesser extent in various stars at non-exploding stages of their evolution. At least it seems likely that the effect can appear in the objects with keV temperatures and densities of a few tens g/cm$^3$ to a few hundreds g/cm$^3$.

It is hardly possible, however, to provide a universal rule by which one can accurately predict the strength of suprathermal effects is stars. It depends on the stellar plasma condition and may differ significantly for different objects. Therefore, each particular case should be studied individually.


\begin{thebibliography}{99}

\bibitem{arno20}
M.~Arnould and S.~Goriely, Prog. Part. Nucl. Phys. \textbf{112}, 103766 (2020).

\bibitem{gori22}
S.~Goriely, Astron. Astrophys. \textbf{658}, A197 (2022).

\bibitem{tati18}
V.~Tatischeff and S.~Gabici, An. Rev. Nuc. Part. Sci. \textbf{68}, 377 (2018).

\bibitem{gori07}
S.~Goriely, Astron. Astrophys. \textbf{466}, 619 (2007).

\bibitem{liu12}
M.-C.~Liu, M.~Chaussidon, G.~Srinivasan, and K.~D.~McKeegan, Astrophys. J. \textbf{761}, 137 (2012).

\bibitem{beau77}
G.~Beaudet and S.~Shaviv, Astrophys. Space Sci. \textbf{51}, 395 (1977).

\bibitem{petr88}
Iu.~V.~Petrov and A.~I.~Shliakhter, Astrophys. J. \textbf{327}, 294 (1988).

\bibitem{shap04}
A.~I.~Shapiro, Astron. Lett. \textbf{30}, 404 (2004).

\bibitem{voro21a}
V.~T.~Voronchev, Phys. Lett. A \textbf{408}, 127491 (2021).

\bibitem{adel98}
E.~G.~Adelberger \emph{et al}., Rev. Mod. Phys. \textbf{70}, 1265 (1998).

\bibitem{voro17a}
V.~T.~Voronchev, Y.~Nakao, and Y.~Watanabe, J. Phys. G \textbf{44}, 045202 (2017).

\bibitem{voro17b}
V.~T.~Voronchev, Y.~Nakao, and Y.~Watanabe, Phys. Rev. C \textbf{96}, 055803 (2017).

\bibitem{voro19}
V.~T.~Voronchev, J. Phys. G \textbf{46}, 065201 (2019).

\bibitem{ayukov}
S.~V.~Ayukov, private communication.

\bibitem{mesa}
B.~Paxton, L.~Bildsten, A.~Dotter, F.~Herwig, P.~Lesaffre, and F.~Timmes, Astrophys. J. Suppl. Ser. \textbf{192}, 3 (2011).

\bibitem{kame86}
G.~Kamelander, Atomkernenergie-Kerntechnik \textbf{48}, 231 (1986).

\bibitem{xu13}
Y.~Xu, K.~Takahashi, S.~Goriely, M.~Arnould, M.~Ohta, and H.~Utsunomiya, Nucl. Phys. A \textbf{918}, 61 (2013).

\bibitem{ilia10}
C.~Iliadis, R.~Longland, A.~E.~Champagne, A.~Coc, and R.~Fitzgerald, Nucl. Phys. A \textbf{841}, 31 (2010).

\bibitem{till04}
D.~R.~Tilley \emph{et al}., Nucl. Phys. A \textbf{745}, 155 (2004).

\bibitem{brun15}
C.~R.~Brune, J.~A.~Caggiano, D.~B.~Sayre, A.~D.~Bacher, G.~M.~Hale, and M.~W.~Paris, Phys. Rev. C \textbf{92}, 014003 (2015).

\bibitem{fowl67}
W.~A.~Fowler, G.~R.~Caughlan, and B.~A.~Zimmerman, Ann. Rev. Astron. Astrophys. \textbf{5}, 525 (1967).

\bibitem{koni15}
A.~J.~Koning \emph{et al}., TENDL-2015: TALYS-based evaluated nuclear data library. Available at https://tendl.web.psi.ch/tendl\verb|_|2015/tendl2015.html

\bibitem{kies79}
W.~E.~Kieser, R.~E.~Azuma, and K.~P.~Jackson, Nucl. Phys. A \textbf{331}, 155 (1979).

\bibitem{terw08}
G.~Terwagne, G.~Genard, M.~Yedji, and G.~G.~Ross, J. Appl. Phys. \textbf{104}, 084909 (2008).

\bibitem{wies10}
M.~Wiescher, J.~G\"{o}rres, E.~Uberseder, G.~Imbriani, and M.~Pignatari, Annu. Rev. Nucl. Part. Sci. \textbf{60}, 381 (2010).

\bibitem{code}
The code is available at http://cococubed.asu.edu/code\_pages/burn\_hydrogen.shtml

\end{thebibliography}
\end{document}